\documentstyle[preprint,psfig,epsfig,graphics,revtex]{aps}

\begin{document}
\draft
\preprint{}
\begin{title}
New Phase Transition in Polymer Solutions
 \\with Multicomponent Solvent
\end{title}

\author{Zh. S. Gevorkian$^{1,2,*}$  and   Chin-Kun Hu$^{1,+}$}

\begin{instit}
$^1$Institute of Physics, Academia Sinica,
Nankang, Taipei 11529, Taiwan\\
$^2$Institute of Radiophysics and Electronics\\
and Yerevan Physics Institute, Alikhanian Br.2\\
375036 Yerevan Armenia
\end{instit}

\begin{abstract}
Considering the density-density correlation function of a concentrated 
polymer solution with multicomponent solvent we find a phase transition
due to the heterogeneity of excluded volume constant. This new phase transition 
implies a strong enhancement of the scattered light intensity in the critical region,
which can explain a recent experiment showing strong light scattering from
a ternary polymer system consisting of polyethylene
oxide (PEO) dissolved in nitroethane and 3methyl-pentane.

\end{abstract}

\vskip 10 cm

\newpage

Physical properties of polymer solutions have attracted much attention 
in recent years \cite{CJ90,pol-sol}. Polymer solutions with multicomponent 
solvent 
are of particular current interest because some polymers require
mixed solvents to reach their $\Theta$-conditions \cite{HF90} and such polymer
solutions have fluctuating excluded volume constants as in the case of
random heteropolymers in a single solvent \cite{SS97}. Therefore,
the investigation of polymer solutions with multicomponent solvent will help
to understand the properties of random heteropolymers, including many 
biopolymers
such as RNA, DNA, protein, etc, which play important roles in functions of
biological systems.

Light scattering is one of the main methods for the investigation of
polymer solutions \cite{MG64}. Particularly, it allows the 
determination of the molecular weight of polymers in dilute solutions
\cite{Deb46}. Many papers have studied the light scattering of
polymer solutions with multicomponent solvents \cite{HY67}.
However, most of these papers considered the scattering from
dilute polymer solutions.
 In the present paper we consider the concentrated polymer 
solutions with  multicomponent solvent. In this case a strong overlapping between the 
polymer chains exists and the excluded-volume interaction between 
the monomers is screened \cite{DE86,HF90}.
 Although the solution is concentrated it still
occupies a very small volume fraction of the system (semidilute
regime).  Such a situation was realized in a recent experiment by To, Kim and Choi (TKC)
\cite{TKC98} who studied light scattering from a ternary polymer system 
consisting of polyethylene oxide (PEO) dissolved in a binary solvent of
nitroethane (NE) and 3methyl-pentane (MP) (PEO/NE/MP) 
and found an abnormal strong light scattering, which does not have any theoretical
explanation. Using the concept of a fluctuating excluded-volume constant 
to calculate the density-density correlation function of a polymer 
solution with a binary solvent, we find a critical behavior for such a 
system when the fraction of one component of the solvent approaches
a critical value. This new phase transition implies 
a strong enhancement of the scattered light intensity in the critical region,
which can explain TKC's abnormal strong light scattering  in PEO/NE/MP \cite{TKC98}.

Consider a solvent the components of which have the same refractive index. 
This means that the solvent will not scatter  light by itself.
However suppose that the components differ from each other by their
ability to dissolve a given homopolymer. Now consider a solution of
concentrated homopolymers in such a multicomponent solvent. Since the
excluded-volume interactions between the monomers depend not only on
themselves but also on their environments, they will fluctuate from point
to point because of the composition fluctuations of the solvent. 
In analogy with the random heteropolymer case \cite{GH01},
the energy of a polymer solution with a fluctuating excluded-volume constant,
in terms of monomer concentration \cite{DE86},
can be written in Gaussian approximation in the form
\begin{equation}
\frac{U}{T}=\frac{1}{2}\int d\vec r c(\vec
r)\left[-\frac{a^2\nabla^2}{12c}+v+v_1(\vec r)\right]c(\vec r).
\label{En}
\end{equation}
Here
\begin{equation}
c(\vec r)=\sum_i\delta(\vec r-\vec R_i)
\label{Con}
\end{equation}
is the microscopic concentration of monomers, 
$c=<c(\vec r)>$ is the average concentration of monomers, $a$ is their
average size and
$v$ is the average value of the excluded volume constant with $v_1(\vec r)$
being its fluctuating part.
The summation in Eq. (\ref{Con}) runs over all chains and monomers 
 and $\vec R_i$ is the coordinate of the $i$-th monomer.
As was mentioned above, the excluded-volume 
constant fluctuations are caused by the composition fluctuations of the
solvent and by the fact that the excluded-volume constant  has different 
 values in different components of the solvent. Therefore we assume that 
the fluctuating part $v_1(\vec r)$ is a Gaussian-distributed random 
function whose correlation function is given by
\begin{equation}
<v_1(\vec r)v_1(\vec {r^\prime})>=B(\vec r-\vec {r^\prime}),\quad <v_1>=0.
\label{exc}
\end{equation}
The Fourier transform of the correlation function $B(\vec r-\vec {r^\prime})$
is given by 
\begin{equation}
B(k)=\frac{w^2}{c}\frac{1}{k^2\xi_s^2+1}.
\label{cfex}
\end{equation}
Here $w$ is the variance (dispersion) of the excluded-volume constant and
$\xi_s$ is the correlation length of the composition fluctuations. 
Note that far away from the critical temperature of the solvent mixture
 $\xi_s$ is very small (less than $2nm$ for NE/MP mixture \cite{CKSA71})
 and fluctuations can be assumed to be almost uncorrelated.
When the solvent consists of only two components, as in the experiment
\cite{TKC98}, the average and the variance of the excluded volume
constant can be estimated as follows
\begin{equation}
v=\Phi v_p+(1-\Phi)v_g, \quad w^2=\Phi(1-\Phi)(v_g-v_p)^2.
\label{var}
\end{equation}
Here $v_p$ and $v_g$ are the excluded volume constants in poor (MP)
and in good (NE) components, respectively; $\Phi$ and $1-\Phi$ are
the fractions of MP and NE in the solvent, respectively.

It is well known that the intensity
of scattered light is proportional to the Fourier transform of
the density-density correlation function of the polymer solution
\cite{MG64}, which is determined as follows
\begin{equation}
G(\vec r,\vec {r^\prime})=\frac{1}{c}\left[<c(\vec r)c(\vec {r^\prime})>-c^2\right].
\label{Cor}
\end{equation}
Here $<....>$ means a thermodynamic average using the energy of
Eq.(\ref{En}). Thus
\begin{equation}
cG(\vec r,\vec {r^\prime})=A\int\prod d\vec R_i [c(\vec r)
c(\vec {r^\prime})- c^2]exp\left(-\frac{U[c]}{T}\right),
\label{Int}
\end{equation}
where $A$ is a normalization constant and the integration goes 
over all monomer coordinates. It is convenient to go to
integration over the monomer concentration $c(\vec r)$ instead of 
$\vec R_i$ in (\ref{Int}) \cite{DE86}. Doing this and calculating the 
Gaussian integral over $c(\vec r)$, we find that the correlation function
$G(\vec{r},\vec{r^\prime})$, representing $vc\xi^2G(\vec{r},\vec{r^\prime})$
above with $\xi$ being $a(12vc)^{-1/2}$, satisfies the equation 
 \cite{GH01}
\begin{equation}
\left[-\nabla^2+\xi^{-2}+\xi^{-2}\frac{v_1(\vec r)}{v}\right]
 G(\vec r,\vec {r^\prime})=\delta(\vec r-\vec {r^\prime}).
\label{Equ}
\end{equation}
The first two terms in (\ref{Equ}) describe the polymer 
solution in a homogeneous solvent and the third one describes the
heterogeneity of the excluded-volume constant.

 It is easy to obtain the bare $(v_1\equiv 0)$ correlation function 
from Eq.(\ref{Equ})
\begin{equation}
G_0(q)=\frac{1}{q^2+\xi^{-2}},
\label{bcf}
\end{equation}
which means that $\xi$ is the correlation length of the polymer
in a homogeneous solution.
For averaging over the realizations of the random field $v_1$, we use the
impurity diagram technique \cite{AGD69}. Expanding Eq.(\ref{Equ})
on $v_1$ and averaging in each order over $v_1$, one can represent
the averaged correlation function in a Dyson form
\begin{equation}
G(q)=\frac{1}{q^2+\xi^{-2}-\Sigma(q)},
\label{dfo}
\end{equation}
where the self-energy $\Sigma(q)$ is determined by the following irreducible diagrams
\begin{equation}
\parbox[c][3.5cm][c]{13cm}{\includegraphics[scale=0.65]{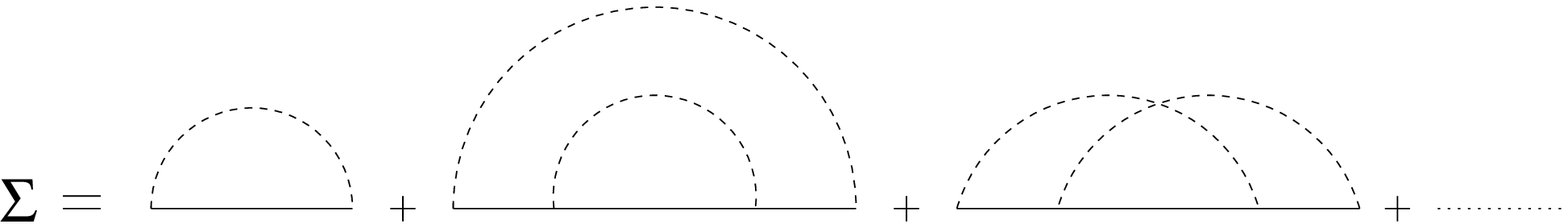}}.
\label{Di}
\end{equation}
The heavy line denotes the bare correlation function of Eq.(\ref{bcf}) and
the dashed line is the Fourier transform of the random field $v_1$
correlation function of Eq.(\ref{exc}), $\xi^{-4}B(k)/v^2$. 
In the leading order we find from Eq.(\ref{Di})
\begin{equation}
\Sigma(\vec q)=\frac{\xi^{-4}}{v^2}\int \frac{d\vec k}{(2\pi)^d}B(k)
G_0(|\vec k-\vec q|).
\label{self}
\end{equation} 
It follows from Eqs.(\ref{dfo})-(\ref{self}) that fluctuations of the
excluded-volume constant lead to renormalization of correlation length.
Substituting Eq.(\ref{bcf}) into Eq.(\ref{self}), one obtains 
from Eq. (\ref{dfo}) the following expression for the correlation function
\begin{equation}
G(q)=\frac{1}{q^2+\xi_R^{-2}},
\label{rcf}
\end{equation}
where the renormalized correlation length is determined as
\begin{equation}
\xi_R^{-2}=\xi^{-2}\left(1-\frac{w^2}{v^2}\frac{1}{4\pi c\xi\xi_s(\xi+\xi_s)}\right).
\label{rcl}
\end{equation}
To obtain Eq.(\ref{rcl}) we take the limit $q\rightarrow 0$ in
Eq.(\ref{self}). This approximation is justified for consideration of 
scattering in the optical region because $\lambda\gg \xi,~\xi_s$ (see below).
Equation (\ref{rcl}) implies that the renormalized correlation length $\xi_R$ 
diverges  when the heterogeneity parameter $g={w^2}/{v^2}$ approaches the
critical value:
\begin{equation}
g_c=4\pi c\xi\xi_s(\xi+\xi_s).
\label{cvh}
\end{equation}
This means that the system undergoes a second order phase transition
at $g_c$ \cite{GH01}. Note that in $\Theta$ conditions, when $v\to 0$, the limit
$g\to g_c$ is reached for any nonzero value of variance $w\ne 0$ because
$g$ behaves as $v^{-2}$. Note that for the binary solvent the critical
value of $\Phi$, $\Phi_c$, can be found from
the equation $g(\Phi_c)=g_c$, where 
$g(\Phi)=w^2(\Phi)/v^2(\Phi)$ is determined by Eq.(\ref{var}).


Now let us present the physical reason for
the increases of density fluctuations and the
renormalized correlation  length $\xi_R$ with the heterogeneity parameter 
$g=w^2/v^2$, which eventually leads to the phase transition.
When $g$ is 0 or very small, to maximize the entropy (minimize the 
free energy) polymer chains prefer to take conformations which make 
the density of monomers in solution almost constant 
with small fluctuations. When $g$ is increasing, the solvent
is more and more heterogeneous consisting of regions with small and
large excluded volume constants. In such cases, the monomers prefer the
regions with small excluded volume constant  
to minimize the excluded volume energy. This leads to large  density
fluctuations of monomers because of the composition fluctuations of the solvent. 
When the heterogeneity parameter $g$ exceeds the critical value $g_c$, 
the trend to minimize the excluded volume energy having a heterogeneous
structure becomes more important than the trend to maximize the entropy having almost
constant density. Since the polymer solution is concentrated, 
the overlapping between the polymer chains exists and the fluctuations
can propagate through polymer chains, which leads to the divergence 
of $\xi_R$ at $g_c$.

Now consider how the properties of polymers in a multicomponent solvent just 
mentioned could be reflected on the light scattering. As was stated above 
the intensity of scattered light is proportional
to $G(q)$ of Eq.(\ref{rcf}). 
From Eqs.(\ref{rcf}), (\ref{rcl}) and (\ref{cvh}) we obtain

\begin{equation}
I(k)\sim \frac{1}{k^2+\xi^{-2}(1-\frac{g(\Phi)}{g_c})}.
\label{ins}
\end{equation}
Here $k=\frac{4\pi}{\lambda}n \sin\theta/2$ with $\lambda$ being the wavelength
of the incident light, $n$ being the refractive index of the medium, and
$\theta$ being the scattering angle. 
Consider the asymptote of Eq.(\ref{ins}). Far away from the critical point
 $g\ll g_c(\Phi\ll \Phi_c)$ the renormalized correlation length is almost
equal to the
 bare one $\xi_R\sim\xi$. In the optical region $\lambda\gg \xi$, neglecting 
the first term in the denominator of Eq.(\ref{ins}), we find
\begin{equation}
I_0(k)\sim\xi^2.
\label{ins1}
\end{equation}
In this case scattered light is isotropic and the intensity is small
 because in a good solvent the correlation
length of a concentrated solution is  of the order of a few nanometers
\cite{HF90}. In the critical region $g\sim g_c$, $\xi_R^{-2}\ll k$
except a small region of angles near $\theta\approx 0$. Therefore
for  scattered light intensity, one has from Eq.(\ref{ins})
\begin{equation}
I(k)\sim \frac{\lambda^2}{16\pi^2 n^2 \sin^2\theta/2}.
\label{ins2}
\end{equation}
It follows from Eq.(\ref{ins2}) that at the critical value $g\sim g_c$
of the heterogeneity parameter  scattered light is strongly anisotropic
and the intensity is much larger than that predicted by Eq.(\ref{ins1})
(see below).

Now we want to apply our theory to polymers dissolved in a binary solvent.
The experimental system consists of polyethylene oxide (PEO)
dissolved in nitroethane+3-methyl-pentane (PEO/NE/MP) where
nitroethane (NE) is a good solvent for PEO and the methyl-penthane (MP) 
is a poor one. The molecular
weight and the concentration of PEO in the solvent 
were $9 \times 10^5$ and $0.075$ mg/cc, respectively. 
Note that a $MW=9\times10^5$ molecular weight PEO solution in a marginal
(mixture of good (NE) and poor(MP)) solvent should be treated as a
concentrated one because its concentration $0.075$ mg/cc significantly
exceeds the overlapping concentration of the chains (see, for example,
\cite{DE86}, \cite{HF90}).
The volume
fraction of MP was controlled in the experiment and the measurements were
done at different values of the fraction. The refractive indices of 
NE and MP are very close to each other, $1.39$ and $1.38$, respectively.
This means that  light scattering caused by composition 
fluctuations of the binary solvent could be significant only near the
critical temperature of the mixture. Note that the binary mixture
NE/MP has an upper critical temperature at $26.5^o$ C and the critical
composition is $\Phi_0=65\%$ MP by volume \cite{WMH69}. An argon ion
laser with the wavelength $\lambda=488nm$ was used in the experiment.
The temperature
of the sample used by TKC was  $42^o$ C , far away from the critical
temperature. Nevertheless when the volume fraction of MP was varied from
$0.58$ to $0.7$ an abrupt increase in the intensity of  scattered 
light was observed.
The experiment shows more than two orders of magnitude increase in the
scattering intensity within the narrow composition range. At higher
composition $\Phi>0.7$ the intensity is greatly reduced (see Fig. 1 
and Sec. 3 of \cite{TKC98}).

Note that when the fraction
of MP is small the average excluded-volume constant, according to 
Eq.(\ref{var}), is mainly determined by the good solvent NE, and the
heterogeneity parameter is small $g\ll g_c$. Therefore scattered
light intensity is determined by Eq.(\ref{ins1}). Increasing the fraction
of MP we increase the heterogeneity parameter which reaches the critical
region $g\sim g_c$. In this case  scattered light intensity is determined by 
Eq.(\ref{ins2}). Comparing the intensities Eqs.(\ref{ins1})
and (\ref{ins2}) at scattering angle $\theta=90^o$ (as in the experiment), we have
\begin{equation}
\frac{I}{I_0}=\left(\frac{\lambda}{2\sqrt{2}\pi \xi n}\right)^2.
\label{fra}
\end{equation}
Substituting the experimental values \cite{TKC98} of
$\lambda=488nm$, $n=1.38$ and $\xi\approx 4nm$ in Eq. (\ref{fra}),
we have $I/I_0\sim 100$.
This is consistent with what was observed in the experiment \cite{TKC98}.
Note that the other characteristic feature of light scattering:
isotropic angular distribution in the good solvent regime and anisotropic
distribution in the critical region predicted by the theory were also observed
in the experiment \cite{TKC98}.

Now let us present a consistency check of some experimental data.
It follows from  Eq. (\ref{var}) that the heterogeneity
parameter $g=w^2/v^2$ depends on $\Phi$ and $x=v_p/v_g$ and can be written
as 
\begin{equation}
g(\Phi,x)=\frac{\Phi(1-\Phi)(1-x)^2}{(x\Phi+1-\Phi)^2}.
\label{het}
\end{equation}
Differentiating Eq. (\ref{het}) with respect to $\Phi$ and setting
the result to 0, we find that the maximum of $g$ is reached
at $\Phi_0=1/(x+1)$ and $g_{max}=(1-x)^2/4x$. Assuming that $x\sim 0.4$, 
one has $g_{max}\sim 0.22$ and $\Phi_0 \sim 0.7$ which is consistent
with the critical fraction of MP: $\Phi\sim 0.6-0.7$.
Substituting from the experiment the values 
$\xi \sim 4nm$, $\xi_s\sim 1nm$ and $c=12\times 10^{17}cc^{-1}$ 
into Eq. (\ref{cvh}), we find that $g_c \sim 0.22$,
which is consistent with the $g_{max}$. 
Please note that from only one phenomenological parameter 
$x\sim 0.4$ we obtain $g_{max}$ and $\Phi_0$ which are
consistent with experimental data. 
 
As mentioned above, the experiment \cite{TKC98} was done at 
temperatures far above the critical temperature of the NE/MP
mixture. The question arises of how the scattered light intensity will behave
when the temperature decreases and approaches the critical temperature
of the solvent. Obviously the correlation length of the composition
fluctuations $\xi_s$ will increase under the temperature decreasing.
Accordingly, from Eqs.(\ref{cvh}) and (\ref{ins}), the 
scattered light intensity will decrease. However, near the critical 
temperature, although the difference between the refractive indices 
is small, the scattering by the solvent itself becomes essential because
of the large value of correlation length $\xi_s$. Therefore the intensity 
of scattered light will first decrease with decreasing temperature, 
then begin increasing near the critical temperature of the NE/MP mixture. 
This behavior was observed in later experiments \cite{To01}.

Note that our calculation here corresponds to a mean field approximation.
 The divergence of the correlation length at the
critical point $\xi_R\sim \xi(g_c-g)^{-\nu}$ with the exponent 
$\nu=1/2$ (see Eq. (\ref{rcl})) is the reflection of this fact. 
However in the vicinity
of critical point $g\sim g_c$ one should take into account large
fluctuations. We have shown \cite{GH01} that effectively these
fluctuations are described by the field-theoretical $O(N)$ model at
$N=0$. Therefore the correct value of critical index $\nu=0.588$ (see,
for example, \cite{ZJ96}) follows from this correspondence. A more
detailed experiment on light scattering in the above-mentioned
ternary system could check this theoretical prediction.

In summary, we have presented a theory of phase transition due to the 
heterogeneity of the excluded-volume constant in a concentrated 
polymer solution with multicomponent solvent. The abnormal scattering
of light observed in the experiment 
of \cite{TKC98} is a manifestation of such critical behavior.
Our theory describes the main characteristic features of
the experiment \cite{TKC98} qualitatively as well as quantitatively.

We would like to thank Jonathan Dushoff for a critical reading of
the manuscript, N. Izmailian and Kiwing To for valuable comments.
This work was supported in part by the National Science Council of the
Republic of China (Taiwan) under Contract No. NSC 90-2112-M-001-074.

\end{document}